\documentclass[aps,prd,twocolumn,nofootinbib,showpacs,superscriptaddress]{revtex4-1}

\usepackage{amsfonts}
\usepackage{amsmath}
\usepackage{amssymb}
\usepackage{bm}
\usepackage{dcolumn}
\usepackage{epsfig}
\usepackage{graphicx}
\usepackage{graphics}
\usepackage[latin1]{inputenc}
\usepackage{latexsym}
\usepackage{rotating}
\usepackage{hyperref}
\usepackage[usenames]{color}
\usepackage{float}
\usepackage{xspace} % Sensible space treatment at end of simple macros
\usepackage{mathrsfs}
\usepackage{subfigure}
\usepackage{enumitem}
\usepackage{ulem}
\definecolor {darkgreen}{rgb}{0.2,0.7,0.2}
\graphicspath{ {/Figs/scalar_charge/} }

\newcommand\be{\begin{equation}}
\newcommand\ba{\begin{eqnarray}}
\newcommand\ee{\end{equation}}
\newcommand\ea{\end{eqnarray}}

\newcommand{\nn}{\nonumber}

\newcommand{\mat}{{\mbox{\tiny mat}}}

\newcommand{\SC}{{\mbox{\tiny sc}}}
\newcommand{\BD}{{\mbox{\tiny BD}}}
\newcommand{\DEF}{{\mbox{\tiny DEF}}}
\newcommand{\J}{{\mbox{\tiny J}}}
\newcommand{\E}{{\mbox{\tiny E}}}

\newcommand{\ppN}{{\mbox{\tiny ppN}}}

\newcommand{\bbn}{{\mbox{\tiny bbn}}}

\begin{document}
\title{The Effect of Cosmological Evolution on Solar System Constraints \\ and on the Scalarization of Neutron Stars in Massless Scalar-Tensor Theories} 

\author{David Anderson}
\affiliation{eXtreme Gravity Institute, Department of Physics, Montana State University, Bozeman, MT 59717, USA.}

\author{Nicol\'as Yunes}
\affiliation{eXtreme Gravity Institute, Department of Physics, Montana State University, Bozeman, MT 59717, USA.}

\author{Enrico Barausse}
\affiliation{CNRS, UMR 7095, Institut d'Astrophysique de Paris, 98bis Bd Arago, 75014 Paris, France}
\affiliation{Sorbonne Universit\'es, UPMC Univ Paris 06, UMR 7095, 98bis Bd Arago, 75014 Paris, France}

\date{\today}

%%%%%%%%%%%%%%%%%%%%%%%%%%%%%%%%%%%%%%%%%%%%%%%%%
\begin{abstract} 

% Intro
Certain scalar-tensor theories of gravity that generalize Jordan-Fierz-Brans-Dicke theory are known to predict non-trivial phenomenology for neutron stars. 
In these theories, first proposed by Damour and Esposito-Far\`ese, the scalar field has a standard kinetic term, and couples conformally to the matter fields. The weak equivalence principle is therefore satisfied, but scalar effects may arise in strong-field regimes, e.g. allowing for violations of the strong equivalence principle in neutron stars (``spontaneous scalarization'') or in sufficiently tight binary neutron-star systems (``dynamical/induced scalarization''). 
The original scalar-tensor theory proposed by Damour and Esposito-Far\`ese is in tension with solar-system constraints (for couplings that lead to scalarization), if one accounts for cosmological evolution of the scalar field and no mass term is included in the action.
We here extend the conformal coupling of that theory, in order to ascertain if, in this way, solar-system tests can be passed, while retaining a non-trivial phenomenology for neutron stars. We find that even with this generalized conformal coupling, it is impossible to construct a theory that passes both Big-Bang nucleosynthesis and solar-system constraints, while simultaneously allowing for scalarization in isolated/binary neutron stars.

\end{abstract}
%04.30.Db Wave generation and sources
% 04.50.Kd Modified theories of gravity
% 04.25.-g Approximation methods; equations of motion
%04.25.Nx Post-Newtonian approximation; perturbation theory; related approximations
%97.60.Jd Neutron stars

\maketitle
%\tableofcontents
\allowdisplaybreaks[4]

%%%%%%%%%%%%%%%%%%%%%%%%%%%%%%%%%%%
\section{Introduction}

%par1: The importance of experimental relativity
The emission of gravitational waves by binary systems of compact objects is a key prediction of General Relativity (GR) and other relativistic gravitational theories~\cite{Yunes:2013dva,Berti:2015itd}. Their existence has been established indirectly (through their backreaction on the orbital motion) by timing the period of pulsar binaries~\cite{Kramer:2016kwa,Wex:2014nva}, and directly by Advanced LIGO~\cite{Collaboration:2016ki,Abbott:2016nmj} through the observation of the coalescence of black-hole binaries. Gravitational waves also allow for exquisite tests of gravitation in extreme gravity regimes~\cite{Yunes:2013dva,Berti:2015itd}, since the binaries of compact objects that most copiously emit them involve strong and dynamical gravitational fields, and highly relativistic velocities. Therefore, these tests are qualitatively different than solar-system experiments~\cite{will-living}, which probe gravity in the quasi-stationary, weak-field regime.

%par2: Why to consider scalar-tensor theories
One of the most natural extensions of GR is the scalar-tensor theory class~\cite{1970PhRvD...1.3209W,1970ApJ...161.1059N,1968IJTP....1...25B}. This class is defined (in the so-called Einstein frame) through the inclusion in the Einstein-Hilbert action of a scalar field $\varphi$ with a canonical kinetic term, a potential and a (conformal) coupling to matter. Different members of this class are defined by the choice of potential and conformal coupling. For example, when the potential is chosen to be zero or to be a canonical mass potential, one obtains the so-called massless or massive scalar-tensor theory subclass. When, in addition, one chooses the logarithm of the conformal coupling to be linear in the scalar field, one obtains massless or massive Jordan-Fierz-Brans-Dicke (JFBD) scalar-tensor theory~\cite{jordan,Jordan-book,fierz,Brans:1961sx}, while when one allows for a quadratic scalar field term (whose magnitude is controlled by a constant $\beta$), one obtains massless or massive Damour-Esposito-Far\`ese (DEF) scalar-tensor theory~\cite{Damour:1993hw,Damour:1992we}.   

%par3: Why is it a natural choice
Scalar-tensor theories are a natural choice of study because they arises in the low-energy limit of more fundamental quantum gravitational theories. For example, scalar-tensor theories arise in heterotic string theory upon 4-dimensional compactification~\cite{polchinski1}, in higher-dimensional theories, such as Kaluza-Klein models~\cite{Duff:269938} and braneworld models~\cite{Randall:389728,1999PhRvL..83.3370R}, and in effective field theories of inflation~\cite{Weinberg:2008hq}. Typically, these theories predict a multitude of scalar fields (not just one), with couplings that include the ones we focus on here, but also encompass more complicated functions of the curvature tensor.

%par4: Constraints on scalar-tensor theories 
Different scalar-tensor theories have been constrained to different degrees with different observations. JFBD theory predicts a modification to the Shapiro time delay of photons propagating on a curved background, which the Cassini probe has verified to be consistent with GR~\cite{Bertotti:2003rm}; such an observation places a very stringent constraint on this theory~\cite{Bertotti:2003rm,2012PhRvD..85f4041A}. The coupling constants of DEF theory, on the other hand, can be tuned so that the theory reproduces exactly GR at first post-Newtonian (PN) order\footnote{The PN approximation is one in which the field equations are expanded in small velocities and weak fields~\cite{Blanchet:2002av}. A term of NPN order is proportional to $(v/c)^{2N}$ with respect to its leading-order (controlling) factor.}, thus evading this constraint, while still allowing for modifications to GR in more ``extreme'' gravity regimes. In particular, non-linear interactions in this theory can lead to the sudden activation of the scalar field in spacetimes containing neutron stars (NS's), isolated or in binaries. 

%par 5: Scalarization in detail
This sudden activation, usually referred to as \textit{scalarization}, is an intrinsically non-linear process that would be a smoking-gun deviation from Einstein's theory. Physically, this occurs when the gravitational binding energy in a matter configuration (e.g. a star) exceeds a certain threshold, which then amplifies the scalar inside matter, even if the asymptotic value of the field at spatial infinity is exponentially small. This phenomenon has been classified into  \emph{spontaneous}, \emph{dynamical} and \emph{induced scalarization}, depending on the specific systems involved. Spontaneous scalarization occurs in isolated and dense compact stars, e.g. neutrons stars, when the compactness of the star exceeds a critical value~\cite{Damour:297525,Damour:1993hw,Damour:1992we,Harada:342978}. Induced scalarization occurs in a neutron-star binary, when one of the stars is exposed to the scalar field of its (already scalarized) companion~\cite{Barausse:2012da,Palenzuela:2013hsa,Shibata:2013pra,Taniguchi:2014fqa}. Dynamical scalarization also occurs in  neutron-star binaries (even ones involving stars that do not spontaneously scalarize in isolation), when the binary's binding energy exceeds a certain critical value~\cite{Barausse:2012da,Palenzuela:2013hsa,Shibata:2013pra,Taniguchi:2014fqa}. 

%Constraining scalarization with GWs
The activation of the scalar field typically produces large deviations from GR in the generation of gravitational waves by NS binaries. The dominant deviation is typically due to the consequent activation of scalar dipolar radiation, which increases the rate at which NS's in binaries spiral into each other. Binary pulsar observations, however, do not see such an increase~\cite{Freire:2012mg}, which then leads to constraints on scalarization. In more detail, binary pulsar observations exclude the presence of
spontaneous scalarization in a given (observed) pulsar mass range, which results in tight constraints on DEF theory~\cite{Freire:2012mg}. Similarly, if the gravitational waves emitted by neutron-star binaries are detected by ground-based interferometers, they may constrain dynamical and induced scalarization~\cite{Sampson:2014qqa}\footnote{These gravitational tests may also be complemented by corresponding ones in the electromagnetic band~\cite{Ponce:2014hha}, if an electromagnetic counterpart is detected.}.

%par5: Cosmological evolution and violation of SS tests
Although DEF theory is a nice playground to explore modifications to GR in the generation of gravitational waves, its massless version at least has a major theoretical problem: {{for choices of the coupling constant $\beta$ that allow for scalarization, the cosmological evolution of the scalar field generically leads to local scalar field values \emph{today} that grossly violate solar-system constraints}}~\cite{Sampson:2014qqa,PhysRevLett.70.2217,PhysRevD.48.3436}. Indeed, the cosmological equation of motion for the scalar field resembles that of a damped oscillator, whose potential is a function of the conformal coupling and the cosmology~\cite{PhysRevLett.70.2217,PhysRevD.48.3436}. The DEF choice of the conformal coupling with $\beta < 0$ makes the potential unbounded from below, leading to a cosmological runaway attractor solution for the scalar field~\cite{Sampson:2014qqa}. The only way to avoid this runaway behavior in massless DEF theory is to choose $\beta > 0$, as shown by Damour and Nordtvedt~\cite{PhysRevLett.70.2217,PhysRevD.48.3436}, but this is precisely the region of coupling parameter space that does not allow for any type of sudden scalarization (unless $\beta$ is very large, i.e.  $\beta\gtrsim 100$~\cite{Mendes:2016fby}). 

%par6: Possible ways to fix the problem: massive fields
In this paper, we investigate whether massless DEF theory can be modified such that 
the cosmological evolution produces local scalar field values that both pass solar-system tests 
and which allow for scalarization in NS systems. For this purpose, we generalize the conformal coupling to include higher-order terms in the scalar field, while keeping $\beta < 0$ in the quadratic scalar field term. Based on the phenomenological oscillator picture discussed above, we must construct a conformal coupling such that the oscillator potential contains a global minimum, allowing the field to settle there at late times. By choosing the higher-order terms such that this is guaranteed, this modified massless DEF theory can pass solar-system constraints upon cosmological evolution, while also (potentially)
 allowing the scalar field to have a non-trivial solution inside NS's\footnote{An alternative approach we did not pursue in this paper is to allow the scalar field to be massive~\cite{Pretorius:2016wp}.}. 

\begin{figure}[h]
	\centering
	\includegraphics[width=3.5in]{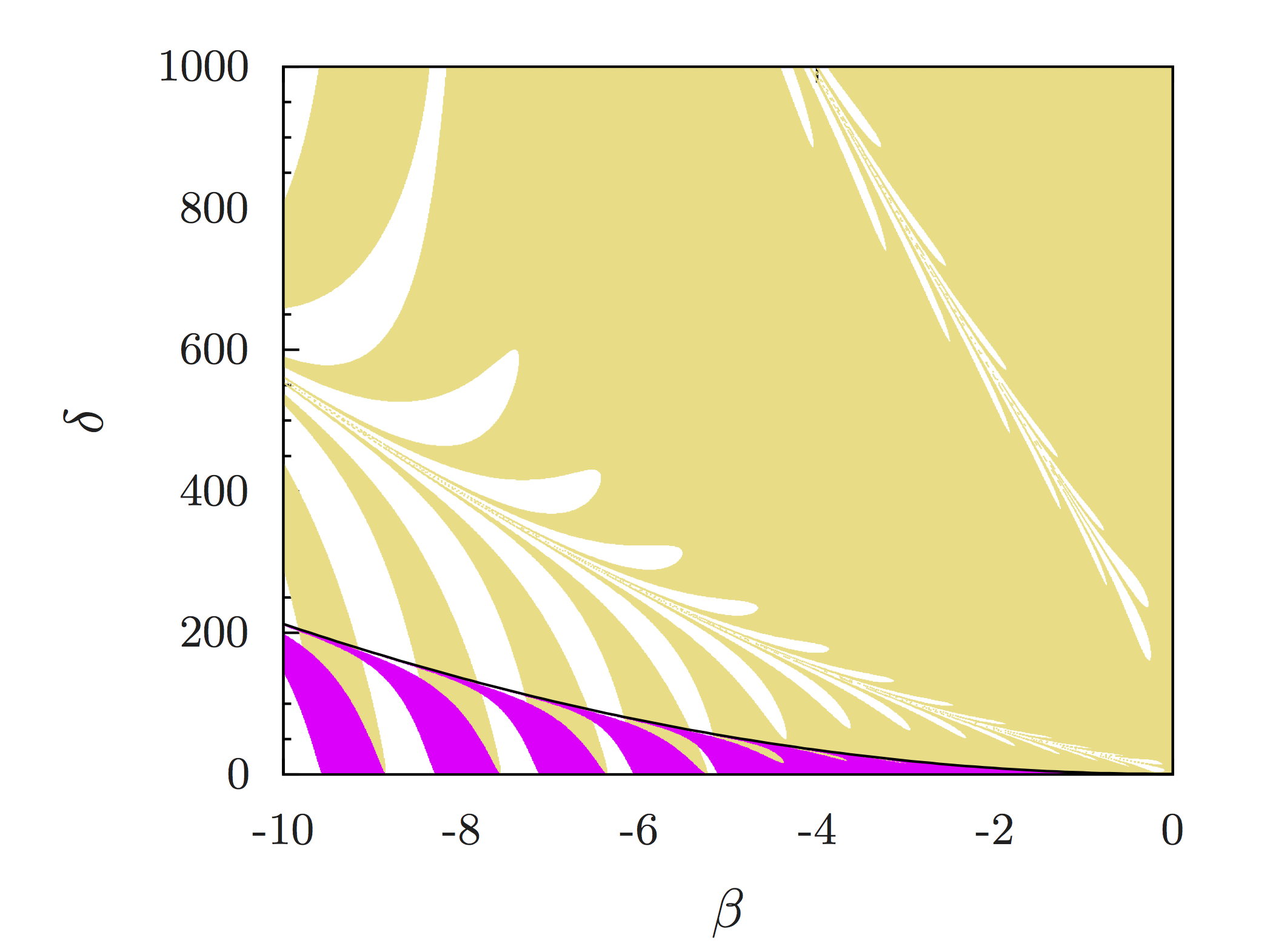}
	\caption{\label{fig:cosmo__3} (Color Online) Regions of parameter space where modified DEF theory passes solar-system constraints, with blank space being regions that fail and yellow/magenta regions corresponding to different initial conditions at the beginning of matter domination. The region near $\delta = 0$ coinciding with DEF theory, while not visible, is ruled out for generic (i.e. not finely tuned) initial conditions during inflation. 
	}
\end{figure}

%par7: How we attacked the problem and what results we got
We begin by considering the simple case of a cubic polynomial for the coupling: $\alpha(\varphi) = \beta\,\varphi + \delta\,\varphi^3$ with $\beta<0$ and $\delta>0$. This conformal coupling leads to a quartic oscillator potential (a ``Mexican hat'' potential) that contains two global minima and one local maximum. We use Big-Bang Nucleosynthesis (BBN) observations to restrict the values of the scalar field upon exiting the radiation era into two initial data sets. We then evolve each set and show that the modified theory evolves in such a way as to pass solar-system tests today provided $(\beta,\delta)$ are in the colored regions shown in Fig.~\ref{fig:cosmo__3}, with the different colors corresponding to the different initial data sets. We note that, while not visible in the figure, the region near $\delta = 0$ is excluded, since then the conformal coupling potential is quadratic and concave down, so there is no minima.
(Indeed, for  $\delta = 0$ one recovers the original DEF theory.) 
 We conclude this exploration by generalizing our argument to a wider class of polynomial coupling functions, showing in particular that conformal couplings that lead to unbounded oscillator potentials [i.e.~those where the highest power of $\varphi$ is odd, corresponding to even powers in $\alpha(\varphi)$] will generically always produce cosmological runaway solutions that violate solar-system constraints. 

With that study in hand, we then investigate whether modified massless DEF theory (with coupling constants that allow the theory to pass solar-system constraints upon cosmological evolution) also leads to spontaneous scalarization in NS's. We numerically construct NS solutions in the modified theory, with non-vanishing asymptotic (at spatial infinity) values of the scalar field ($\varphi_{\infty}$), as predicted by our cosmological evolutions at the present time. We find that spontaneous scalarization is not present in NS's in the modified theory. We have checked that if we artificially choose $\varphi_{\infty} = 0$, thus neglecting the cosmological evolution of the scalar field, then spontaneous scalarization is present.   

%par9: structure paragraph
The rest of this paper will provide the details of the calculation described above. Section~\ref{The Basics of Scalar-Tensor Theories} introduces the basics of scalar-tensor theories and the constraints solar-system tests place on them. Section~\ref{Cosmological Evolution and solar system Constraints} covers the cosmological evolution of the scalar field in these theories and how we can constrain the $\delta$ parameter we introduce by using solar-system constraints. Section~\ref{Neutron Stars and Scalarization} discusses spontaneous scalarization in NS's and the results we arrive at using the allowed values of ($\beta,\,\delta$) found in Sec.~\ref{Cosmological Evolution and solar system Constraints}. Finally, Section~\ref{Conclusion} concludes by summarizing our results and discussing future work. 

%par10: conventions paragraph
In the remainder of this paper, we use units in which $c=1$. We also follow the conventions of~\cite{MTW}, where Greek letters in index lists stand for spacetime indices.  Other conventions and notation will be defined throughout the paper as they are introduced. 

%%%%%%%%%%%%%%%%%%%%%%%%%%%%%%%%%%%
\section{The Basics of Scalar-Tensor Theories}\label{The Basics of Scalar-Tensor Theories}

This section presents the class of theories we investigate in detail and establishes notation, following mostly~\cite{PhysRevD.48.3436}. The scalar-tensor theory class we consider in this paper can be defined by the action $S_{\J} = S_{\J,g} + S_{\J,\mat}[\chi,g_{\mu \nu}]$,
where the gravitational part is 
\begin{align}
S_{\J,g} &= \int d^{4}x \frac{\sqrt{-g}}{2 \kappa} \left[ \phi R - \frac{\omega(\phi)}{\phi} \partial_{\mu} \phi \partial^{\mu} \phi \right]\,,
\end{align}
with $g$ and $R$ being the determinant and the Ricci scalar associated with the \textit{Jordan-frame} metric $g_{\mu \nu}$, $\omega(\phi)$ is a kinetic coupling function for the \emph{massless} scalar field $\phi$, and $\kappa = 8 \pi G$, with $G$ the (bare) gravitational coupling constant. The matter action $S_{\J,\mat}[\chi,g_{\mu \nu}]$ is a functional of the matter fields $\chi$, which couple directly and only to the metric tensor $g_{\mu \nu}$. The latter implies that these scalar-tensor theories are \emph{metric theories} of gravity. 
  
One can rewrite the above action in a form reminiscent of the Einstein-Hilbert action through a conformal transformation. Let us then consider $g_{\mu \nu} = A(\varphi)^{2} g_{\mu \nu}^{*}$, where we will refer to the conformal metric $g^{*}_{\mu \nu}$ as the \emph{Einstein-frame} metric. 
 With an appropriate choice of the conformal coupling $A$ and by suitably defining the new scalar field $\varphi$, the action becomes
\begin{align}
S_{\E}&= \int d^4x \dfrac{\sqrt{-g_*}}{2\kappa} \left[R_* - 2 g_*^{\mu\nu}\partial_{\mu}\varphi\partial_{\nu}\varphi \right]
\nn \\
&+S_{\E,\mat}[\chi,A^2(\varphi)g^*_{\mu\nu}],
\label{EinsteinAction}
\end{align}
where $g_{*}$ and $R_{*}$ are the determinant and the Ricci scalar associated with the Einstein metric $g_{\mu \nu}^{*}$. In more detail, the transformation between the Jordan and Einstein frames is given by
\be
	A^{2}(\varphi) = \phi^{-1}\,,
\ee
and 
\be
	\alpha(\varphi)^2 \equiv \left(\frac{d \ln A(\varphi)}{d \varphi}\right)^2 = [3 + 2 \omega(\phi)]^{-1}\,,
\label{eq:alpha}
\ee
the latter providing implicitly the relation between $\phi$ and $\varphi$.
For later convenience we define $\alpha=\partial V_{\alpha}/\partial \varphi$ such that
\be
V_{\alpha} = \ln A(\varphi)\,\,,
\label{eq:coupling_potential}
\ee 
which plays the role of a conformal coupling potential in which the scalar field evolves cosmologically. Different choices of $\omega(\phi)$, or equivalently, different choices of $A(\varphi)$ define different members of this massless scalar-tensor class of theories. 

Variation of the Einstein-frame action yields the Einstein-frame field equations
\begin{align}
R_{\mu \nu}^{*} &= \kappa \left(T_{\mu \nu}^{*} - \frac{1}{2} g_{\mu \nu}^* T^{*}\right)\,\,,
\label{eq:Einstein_eqns}
\\
\square_{*} \varphi &= - \dfrac{\kappa}{2}\alpha(\varphi) T_{\mat,*}\,\,,
\label{eq:scalar-field-evol}
\end{align}
where $\square_*$ is the Einstein-frame covariant wave operator, and 
$T^{\mat,*}$ is the trace of the Einstein-frame matter stress-energy tensor $T_{\mu \nu}^{\mat,*}$. The latter is defined through
\be
	T_{\mu\nu}^{\mat,*} \equiv \dfrac{2}{\sqrt{-g}}\dfrac{\delta S_{\E,\mat}[\chi,A^2(\varphi)g^*_{\mu\nu}]}{\delta g_{\mu\nu}^*}\,\,,
\label{eq:stress_energy_mat}
\ee
and is therefore related to the Jordan-frame  stress-energy tensor via $T^{\mu \nu}_{\mat,*} = A^{6} T^{\mu \nu}_{\mat}$. 
The field equations also depend on the total stress-energy tensor, which is simply the sum of the matter and scalar field stress-energy tensors, namely
\begin{align}
T_{\mu \nu}^{*} = T_{\mu \nu}^{\mat,*} + T_{\mu \nu}^{\varphi,*}\,, 
\end{align}
where 
\be
\kappa\,T_{\mu \nu}^{\varphi,*} = 
2\left(\partial_{\mu} \varphi\right) \left(\partial_{\nu} \varphi\right)
- g_{\mu \nu}^* \left(\partial_{\sigma} \varphi\right) \left(\partial^{\sigma} \varphi\right)
\,\,.
\ee
Our study (both cosmological and in NS's) adopts a perfect fluid representation of matter in the Jordan frame. In the Einstein frame, we would like to write
\be
T^{\mu\nu}_{\mat,*} = (\rho_* + p_*)u^{\mu}_*u^{\nu}_* + p_* g^{\mu\nu}_*\,\,,
\label{eq:perfect_fluid}
\ee
where $\rho_*$ is the density, $p_*$ is the pressure, and  $u_*^{\mu}$ is the fluid four-velocity in the Einstein frame. Using the normalization of the four-velocity $(g_{\mu\nu}^* u^{\mu}_* u^{\nu}_* = -1 = g_{\mu\nu} u^{\mu} u^{\nu})$ one finds the relation $u^{\mu}_* = Au^{\mu}$ between Jordan- and Einstein-frame velocities. This result, combined with $T^{\mu \nu}_{*} = A^{6} T^{\mu \nu}$, leads one directly to the relations $\rho_* = A^4 \rho$ and $p_* = A^4 p$ between Jordan- and Einstein-frame quantities.

Different choices of $A(\varphi)$ lead to different functionals $\alpha(\varphi)$, which in turn define different types of scalar-tensor theories. For example, one of the most well-known scalar-tensor theories is JFBD gravity~\cite{jordan,fierz,Brans:1961sx}, in which the conformal coupling takes the form $A_{\BD}(\varphi) = \exp(\alpha_{\BD} \varphi)$, such that $\omega_{\BD}(\phi) = {\rm{const}}$, and thus $\alpha_{\BD}(\varphi) = {\rm{const}} = [1/(3 + 2 \omega_{\BD})]^{1/2}$. Another example is DEF gravity~\cite{Damour:1992we}, defined by the conformal coupling $A_{\DEF}(\varphi) = \exp(\beta_{\DEF} \varphi^{2}/2)$, such that $\alpha_{\DEF}(\varphi) = \beta_{\DEF} \varphi$, with $\beta_{\DEF}$ a constant\footnote{\label{footnote1}~Technically, the theory introduced by DEF also includes a $\varphi$-independent term in $\alpha$, just like in Brans-Dicke theory. However, that term can be set to zero without loss of generality, if one allows the scalar field $\varphi$ to take asymptotically non-zero values far away from a system, c.f. discussion in \cite{Palenzuela:2013hsa,2007arXiv0704.0749D}.}. This theory has attracted considerable attention in recent years, since it has been shown to lead to the excitation of a scalar field near sources with strong gravity~\cite{Damour:1993hw,Barausse:2012da,Palenzuela:2013hsa,Shibata:2013pra,Taniguchi:2014fqa,Sampson:2014qqa} when $\beta_{\DEF} < 0$, without exciting the scalar in the Solar System.  

The functions $A(\varphi)$ and $\alpha(\varphi)$ play a critical role in tests of scalar-tensor theories, because they define the local value of Newton's gravitational constant $G_{N}$ (entering, for example, Newton's Second Law) 
\be
	G_N = G [A^2\,(1 + \alpha^2)]_{\varphi_0}\,\,,
\label{eq:GN}
\ee
and the values of the parameters of the parameterized post-Newtonian (ppN) framework~\cite{1972ApJ...177..757W,1972ApJ...177..775N}. For example, the $\gamma_{\ppN}$ parameter, a measure of how much spatial curvature is produced by a unit rest mass~\cite{TEGP,will-living}, is given in (massless) scalar-tensor theories  by
\begin{align}
	\gamma_{\ppN} - 1 &= - \left.\frac{2 \alpha^{2}}{1 + \alpha^{2}}\right|_{\varphi_{0}}\,,
	\label{eq:gamma_ppn}
\end{align}
where the above expression is to be evaluated at today's value of $\varphi$; similar expressions hold for the other ppN parameters. All ppN parameters have been very well-constrained by solar-system experiments, and in particular, the most stringent constraint on $\gamma_{\ppN}$, $|\gamma_{\ppN} - 1| < 2.3 \times 10^{-5}$ \cite{will-living}, was placed through a verification of the Shapiro time delay of signals from the Cassini spacecraft~\cite{Bertotti:2003rm}. 

%------------------------------------------------------------------------------------------------
\section{Cosmological Evolution and Solar-System Constraints}\label{Cosmological Evolution and solar system Constraints}

In order to determine today's value of $\alpha(\varphi_0)$, we must first understand its cosmological evolution. Consider then the Einstein-frame field equations with a spatially flat Friedmann-Roberston-Walker (FRW) metric with the Einstein-frame scale factor $a_{*}$: 
\begin{align}
3 H_{*}^{2} &= \kappa \rho_* + \dot{\varphi}^{2} \,\,,
\label{eq_for_bbn}
\\
-3 \frac{\ddot{a}_{*}}{a_{*}} &= \frac{\kappa}{2} \rho_* \left(1 + 3 \lambda\right) + 2 \dot{\varphi}^{2}\,\,,
\\
\ddot{\varphi} + 3 H_{*} \dot{\varphi} &= -\dfrac{\kappa}{2} \alpha \rho_* (1 - 3 \lambda) \,\,,
\label{eq:main_evolve}
\end{align}
where $H_{*} = \dot{a}_{*}/a_{*}$ is the Einstein-frame Hubble parameter, the overhead dots stand for derivatives with respect to the Einstein-frame coordinate time $t_*$, $p_*$ and $\rho_*$ are the total
pressure and density (in the Einstein frame) of all the components of the Universe (matter, radiation, dark energy, inflation, etc)
and $\lambda = p_*/\rho_*$ is the usual cosmological equation of state (EoS) parameter. Note that we assume that the energy density and pressure of $\varphi$
are always negligible with respect to those of the other cosmological components, \textit{i.e.} $\varphi$ should not be interpreted as the inflaton or dark energy.
In the matter epoch $\lambda\approx 0$, in the radiation epoch $\lambda \approx 1/3$, and $\lambda \approx -1$ during inflation or after the onset of 
dark energy domination. By introducing a new time coordinate $d\tau \equiv H_* dt_*$, the scalar field satisfies the following evolution equation:
\be
\frac{2}{3-{\varphi}'^2} {\varphi}''+(1-\lambda) {\varphi}'=-(1-3\lambda) {\alpha}({\varphi})
\label{scalarevo}
\ee
where primes denote differentiation with respect to $\tau$.

To gain a better understanding of how $\tau$ varies over timescales we are familiar with, let us look at two cases: the time elapsed since the end of the radiation era until today and the time since the birth of GR (1915) until today. A difference in time $\Delta \tau$ corresponds to 
\be
	\Delta\tau = \ln(a_f^*/a_i^*) = \ln\left(\dfrac{1 + Z_i}{1 + Z_f}\right) +  (V_{\alpha,i} - V_{\alpha,f})\,\,,
	\label{eq:gen_tau_time}
\ee 
where $Z_i$ and $Z_f$ are the initial and final redshifts of a photon traveling a look-back time $\Delta t$ corresponding to $\Delta \tau$. As we will explain later, the terms involving the conformal coupling potential are typically negligible, which allows the analysis to remain independent of the choice of scalar-tensor theory. Therefore, the redshift corresponding to the end of radiation domination is $\sim 3600$ $\Lambda$CDM model, thus, the $\tau$-time that has passed since then is $\Delta\tau \approx 8.2$. From the lookback time, the redshift since 1915 is $7\times 10^{-9}$, and thus $\Delta\tau \approx 7\times 10^{-9}$. Thus, $\Delta\tau \sim 10^{-1}$ actually corresponds to a significant amount of look-back time, $t\sim10^9$ years from today. 

%------------------------------------------------------------------------------------------------

Whether scalar-tensor theories satisfy solar-system constraints today depends on the functional form of $\alpha(\varphi)$ and on the cosmological evolution of the scalar field. The latter resembles the evolution of an oscillator with the velocity-dependent mass $2/(3-\varphi'^2)$, the friction-like term $(1-\lambda)\varphi'$, and the forcing term proportional to $\alpha(\varphi)$. In DEF theory, Damour and Nordtvedt have shown that during the matter-dominated era, $\varphi$ is driven exponentially to zero when $\beta_{\DEF}>0$, such that $\gamma_{\ppN}$ approaches unity at late times~\cite{PhysRevLett.70.2217,PhysRevD.48.3436}. However, Ref.~\cite{Sampson:2014qqa} (see also Refs.~\cite{PhysRevLett.70.2217,PhysRevD.48.3436}) has shown that when $\beta_{\DEF} < 0$ the opposite occurs: $\varphi$ has a \emph{linear} run-away attractor solution that approaches a limiting velocity $\varphi'=\sqrt{3}$. Such an evolution forces $\gamma_{\ppN}-1$ in Eq.~(\ref{eq:gamma_ppn}) to approach $-2$ at late times in the matter era, a value clearly in conflict with solar-system experiments. However, it is precisely when $\beta_{\DEF} < 0$ that strong-field, non-linear effects become important inside NS's and allow for spontaneous/dynamical/induced scalarization, which in turn could lead to clear signatures of deviations from GR in astrophysical observations. Thus, \emph{DEF theory is already ruled out by solar-system observations in the $\beta_{\DEF}$ range of interest ($\beta_{\DEF} < 0$)}~\cite{PhysRevLett.70.2217,PhysRevD.48.3436,Sampson:2014qqa}.

In this paper, we want to investigate whether one can relax the assumption of quadratic conformal coupling such that scalarization continues to occur, yet the theory passes solar-system constraints. Let us then begin by noting that the forcing term in Eq.~(\ref{scalarevo}) can be written as the gradient of a potential, namely the one given in Eq.~\ref{eq:coupling_potential}. In the particle analogy described above, the scalar field starts its evolution in this potential with some initial velocity and position. One then expects that over time this particle will settle to a minimum, if one exists, provided the scalar does not reach the limiting velocity $\varphi'\approx\sqrt{3}$, which effaces the effect of the potential and leads to a run-away attractor solution. Finding a conformal coupling that allows the modified DEF theory to pass solar-system constraints then reduces to finding the appropriate choice of the conformal coupling potential $V_{\alpha}$ (one, in particular, that has a global minimum). 

A word of caution, however, is due before proceeding. The above discussion depends somewhat on the choice of initial conditions for the evolution of the scalar field. Even with a conformal coupling potential that possesses a global minimum, not all initial conditions will lead to scalar field values today that pass solar-system constraints. This is simply because some initial conditions can be so close to the run-away attractor solution that they cannot escape its attraction. We will show below, however, that it is possible to construct a general class of DEF-like theories that can evade solar-system constraints after cosmological evolution for a large set of initial conditions, provided the conformal coupling is chosen appropriately.

%---------------------------------------------------------------------
\subsection{Inflation and Radiation}
\label{sec:inflation}

During inflation ($\lambda = -1$), Eq.~(\ref{scalarevo}) becomes
\be
	\dfrac{1}{3 - \varphi'^2}\varphi'' + \varphi' = -2\alpha(\varphi)\,\,,
\label{eq:inflation_evo}
\ee
which still describes a damped oscillator with a forcing term. In order to study the evolution of the scalar during this era, one needs to prescribe initial conditions at the beginning of inflation. Since the latter are unknown, we will follow Damour and Nordtvedt~\cite{PhysRevLett.70.2217,PhysRevD.48.3436} and take a qualitative approach. Regardless of what the initial conditions are, there are only three possible outcomes upon leaving inflation: 
\begin{enumerate}
\item the scalar can reach its terminal velocity $\varphi'=\sqrt{3}$, and therefore get caught by the attractor solution found in~\cite{Sampson:2014qqa}, 
\item the scalar can end up near (but not necessarily at) a minimum of the potential, 
\item the scalar can be in an intermediate solution (e.g. it may still be rolling down the potential).
\end{enumerate} 
The first possibility leads to theories that never pass solar-system tests today because, once the attractor solution is reached, all subsequent evolution remains on the attractor, regardless of $\lambda$ or $\alpha(\varphi)$. The third outcome is possible in principle, but we find that it requires fine-tuning of the initial conditions, because the friction term efficiently damps any evolution far from the attractor in a short timescale. The second possibility is then the only that remains, and we thus adopt it henceforth to study how the scalar evolves into other cosmological eras. 

Let us now consider the evolution of the scalar during the radiation-dominated era ($\lambda \approx 1/3$). In this era, the forcing term is suppressed, and Eq.~(\ref{scalarevo}) becomes
\be
	\dfrac{2}{3 - \varphi'^2}\varphi'' + \dfrac{2}{3}\varphi'  = 0\,\,,
\label{eq:rad_evo}
\ee
which notice is completely independent of the conformal coupling. Damour and Nordtvedt showed~\cite{PhysRevLett.70.2217,PhysRevD.48.3436} that, during the radiation era, the scalar field evolves according to 
\be
\varphi(\tau) = \varphi_{r} - \sqrt{3}\ln\left[K e^{-\tau} + (1+ K^2 e^{-2\tau})^{1/2}\right]\,\,,
\label{eq:scalar_radiation}
\ee
with
\be
K = \dfrac{\varphi_i' \sqrt{3}}{\sqrt{1 - \varphi_i'^2/3}}\,\,,
\ee
where $ \varphi_r$ is a constant and $\varphi_i'$ is the particle's velocity upon leaving inflation. As long as the latter does not approach the limiting value of $\sqrt{3}$ of the run-away attractor solution (to prevent $K$ from approaching infinity), then Eq.~(\ref{eq:scalar_radiation}) tells us that the velocity at the end of radiation domination will be damped away.  This can be seen by considering the amount of $\tau$-time that elapses during the radiation-dominated era: the $\tau$-time between the end of the radiation-dominated era (0.75 eV, $Z \approx 3600$) and the electroweak (EW) phase transition (100 GeV, $Z \approx 10^{15})$, the QCD phase transition (150 MeV, $Z \approx 10^{12})$, or electron/positron pair ($e^-e^+$) annihilation (500 keV, $Z \approx 10^9)$ is $\tau_{EW} = 25.6$, $\tau_{QCD} = 19.1$, and $\tau_{e^-e^+} = 13.4$ respectively [from Eq.~\eqref{eq:gen_tau_time}]. Even the shortest of these $\tau$-times is long enough to allow the scalar velocity to become exponentially damped by the end of the radiation era.

The evolution in the radiation era also determines the end position of the scalar in the conformal coupling potential at the beginning of the matter-domination era. Let us investigate this by considering the constraint on the gravitational constant from BBN, which took place at temperatures between 10 and 0.1 MeV in the radiation era. The speed-up factor $\xi_{\bbn} := H/H_{GR}$ quantifies deviations of the expansion rate from the GR prediction, caused by changes to the (standard) gravitational constant. Here, $H^2 = (8\pi/3)GA^2_R\rho_R$ is the observed expansion rate in scalar-tensor theory, where $\rho_R$ and $A_R$ are the Jordan-frame energy density and the conformal coupling during the radiation era [this can be derived from Eq.~(\ref{eq_for_bbn}) with $\dot{\varphi}=0$]. On the other hand, $H_{GR}^2 = (8\pi/3)G_N\rho_R$ is the expansion rate predicted by GR, where $G_N$ is the (standard) gravitational constant we measure today given in Eq.~(\ref{eq:GN}). The speed-up factor then becomes
\be
\xi_{\bbn} = \left(\dfrac{H}{H_{GR}}\right) = \left(\dfrac{G A_R^2}{G_N}\right)^{1/2} = \dfrac{1}{\sqrt{1+\alpha_0^2}}\dfrac{A_R}{A_0}\,\,,
\label{bbn_condition}
\ee
where the $R$ and $0$ subscripts represent values at the end of the radiation era and the present values, respectively. Current tests relating the speed-up factor to the abundance of Helium \cite{2011LRR....14....2U} tell us that
\be
	|1 - \xi_{\bbn}| \leq \frac{1}{8}\,\,,
\label{bbn_constraint}
\ee
and solar-system tests limit $\alpha_0^2$ to be $\lesssim 10^{-5}$. This leads to $\xi_{\bbn}\sim A_R/A_0$, which through Eq.~(\ref{bbn_constraint}) leads to a constraint on $A_R$ given by
\be
	\left|1 - \dfrac{A_R}{A_0}\right| \leq \frac{1}{8}\,\,.
\label{bbn_constraint_ratio}
\ee
The largest deviations from the predictions of GR will be achieved by saturating this constraint, such that $A_{R}/A_{0} = 7/8$ or $A_{R}/A_{0} = 9/8$.  Thus, using Eq.~(\ref{eq:coupling_potential}), we arrive at the requirement
\be
	 V_{\alpha,0} + \ln(7/8) \,\leq\, V_{\alpha,R} \,\leq\, V_{\alpha,0} + \ln(9/8)\,\,.
\label{potential_rad}
\ee
In what follows, we will use these BBN-compatibility condition to determine the initial conditions
for the evolution of the scalar field at the beginning of the matter-dominated era. 

%--------------------------------------------------------------------
\subsection{Matter and Dark Energy Domination}\label{matter_sec}
%--------------------------------------------------------------------

Let us now solve for the evolution of the scalar field during the matter-dominated era ($\lambda = 0$), during which Eq.~(\ref{scalarevo}) reduces to
\be
	\dfrac{2}{3 - \varphi'^2}\varphi'' + \varphi' = -\alpha(\varphi)\,\,.
\label{eq:matter_evo}
\ee
Let us consider a generic scalar-tensor theory defined by
\be
	{\alpha}(\varphi)=\sum\limits_{n=1}^{\infty} a_n\varphi^n = \beta \varphi + \gamma \; \varphi^{2} + \delta \; \varphi^{3} + \ldots \,,
	\label{eq:gen_alpha}
\ee
where we have neglected the $n=0$ term associated with JFBD gravity (c.f.~footnote \ref{footnote1}).  For consistency with  standard DEF theory, we define $a_{1} = \beta$, and for later convenience, we define $a_{2} = \gamma$ and $a_{3} = \delta$.  As we will show later, the $\gamma$ term leads to a modified DEF theory that typically violates solar-system constraints, so let us ignore it for now. We then focus first on modified DEF theories defined by 
\ba
	\alpha(\varphi) &=& \beta\,\varphi + \delta\,\varphi^3\,\,,
	\label{alpha_delta}
	\\
	V_{\alpha}(\varphi) &=& \frac{\beta}{2}\varphi^2 + \frac{\delta}{4}\varphi^4\,\,,
	\label{eq:coupling_pot_delta}
\ea
where we will take $\beta<0$ and $\delta>0$. (We will consider $\alpha(\varphi)$ with higher order $\varphi$ terms later.) One recognizes that the conformal coupling potential presents a  ``Mexican-hat'' shape, with negative curvature near the origin and two global minima that prevent the scalar from running off to infinity. For appropriate initial conditions and values of $\delta$, one expects the scalar field to execute damped oscillatory motion about the global minima, without ever reaching the terminal velocity and the attractor solution, and eventually settling down near one of the minima by today.

%--------------------------------------------------------------------
\subsubsection{Initial Conditions}

\begin{figure}[h]
	\centering
	\includegraphics[width=3.5in]{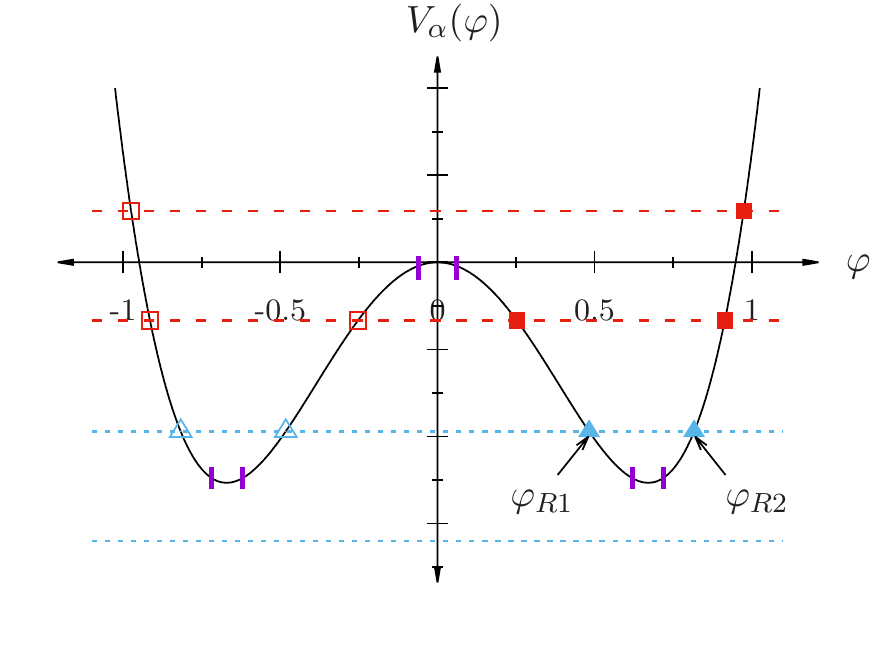}
	\caption{\label{fig:initial_conditions} (Color Online) Schematic diagram of the conformal coupling  potential, where we used $\beta = -4.5$ and $\delta = 10$. Purple dashes represent boundaries of the regions consistent with ppN constraints (determined by Eq.~(\ref{eq:gamma_ppn})). Horizontal dashed lines (as well as the squares and triangles marking their intersections with the potential) are a representation of BBN constraints on the speed-up factor and determine the regions in the potential consistent with nucleosynthesis at the end of the radiation era. Because of the symmetry of the potential, we shade in the {squares/triangles} on the right  to indicate that one only needs to consider these initial conditions without loss of generality. $\varphi_{R1}$ and $\varphi_{R2}$ indicate the points we investigate in this paper.
}
\end{figure}

Before solving Eq.~(\ref{eq:matter_evo}) we must first quantitatively determine the initial conditions for the scalar field evolution at the beginning of the matter era. To find these, we first determine where the scalar field could be today in the conformal coupling potential if the theory is to pass solar-system constraints, and then use the BBN constraint condition to determine where the field had to be at the beginning of the matter era. With our choice of $\alpha(\varphi)$ in Eq.~(\ref{alpha_delta}) and noting the $\alpha(\varphi)^2$-dependence in Eq.~(\ref{eq:gamma_ppn}), it is then clear that there are six possible values of $\varphi$ today that saturate the Cassini bound $1 - \gamma_{\ppN} \leq 2.3\times 10^{-5}$. These six values of $\varphi$ are indicated by purple dashes in Fig.~\ref{fig:initial_conditions} near the extrema of the potential.

The next step requires that we apply the BBN constraint derived in Eq.~(\ref{potential_rad}) to map these possible values of $\varphi$ today to possible initial values of the scalar field at the beginning of the matter era. Equation~(\ref{potential_rad}) tells us that, at the very most, the scalar field can sit no more than $\ln(7/8)$ below or $\ln(9/8)$ above where it sits in the conformal coupling potential today; this is indicated by the blue and red horizontal dashed lines in Fig.~\ref{fig:initial_conditions}. In particular, the red dashed lines correspond to applying the BBN constraint to values of $V_{\alpha,0}$ that lie near zero, while the blue dashed lines were used for those that lie near the minimum of the potential. Figure~{\ref{fig:initial_conditions}} clearly shows that there are 10 possible initial conditions for the scalar field at the beginning of matter domination that can potentially lead to scalar field configurations that satisfy solar-system constraints today.

Not all of these initial conditions are physically well-motivated based on the previous arguments we presented. We have previously argued that inflation leaves the scalar field near the minimum of the potential and the radiation era effectively keeps it there, since the scalar velocity is damped away completely. The initial positions lying near $V(\varphi) = 0$ (red squares) in Fig.~\ref{fig:initial_conditions} are inconsistent with these physical arguments, since they \emph{do not} lie near the minimum, and thus, they will be neglected in what follows. We now only need to consider initial positions near the minimum of the potential (blue triangles), and because of the symmetry of the potential we need only consider one of the two sets; the evolution of the scalar field that starts at the solid blue triangles will be identical to that which starts at the empty blue triangles, and thus, leads to the same conclusions.

The initial positions labeled $\varphi_{R1}$ and $\varphi_{R2}$ in Fig.~\ref{fig:initial_conditions} with initial velocity $\varphi' = 0$ at the end of the radiation era are the initial states of the scalar field we aim to investigate. Typically, one need to consider both, but for a sufficiently large $\delta$ (relative to $\beta$) only $\varphi_{R2}$ exists. This is because when $\delta \gg \beta$, the conformal coupling potential becomes very shallow and the top blue line can be above the extremum at $V_{\alpha}(\varphi) = 0$, leading only to $\varphi_{R2}$ (the other initial condition $\varphi_{R1}$ becomes imaginary).

%--------------------------------------------------------------------
\subsubsection{Cosmological Evolution}

The evolution of the scalar field during matter domination, as given in Eq.~(\ref{eq:matter_evo}), is that of a damped oscillator. Provided the attractor solution is not reached, i.e.~provided $\varphi'$ does not reach its limiting value of $\sqrt{3}$, then the scalar will exhibit damped oscillatory motion in the potential. For solar-system tests to be passed, then, one needs $\alpha(\varphi_{0})$ to be small enough after a time $\tau_{0}$ has elapsed from the beginning of the matter-dominated era.  

Let us begin by calculating what this $\tau_{0}$ must be. Recalling that $d\tau = H_* dt_*$, one has that $\tau = \ln a_* + \text{const}$. If we set $\tau$ to zero at the end of the radiation era and recall that $a = A(\varphi)a_*$ and $V_{\alpha}=\ln A(\varphi)$, today corresponds to
\be
	\tau_0 = \ln a_{*,0} - \ln a_{*,R} = \ln(1+Z_R) + (V_{\alpha,R} - V_{\alpha,0})\,\,,
\ee
where $Z_R\approx 3600$  is the redshift of the end of the radiation era. For an evolution that satisfies solar-system constraints, the particle must settle toward a minimum of $V_{\alpha}$, which means the last term above will always be a positive number. The most stringent constraints on these theories arise when we neglect this last term and demand that solar-system constraints be satisfied at least by $\tau_0 = \ln(1+ Z_R) \approx 8.2$. The inclusion of the last term would make $\tau_0$ larger, which would then allow the scalar field more time to settle near the minimum of $V_{\alpha}$, thus leading to weaker restrictions on $\delta$. 

\begin{figure}[h]
	\centering
	\includegraphics[width=3.5in]{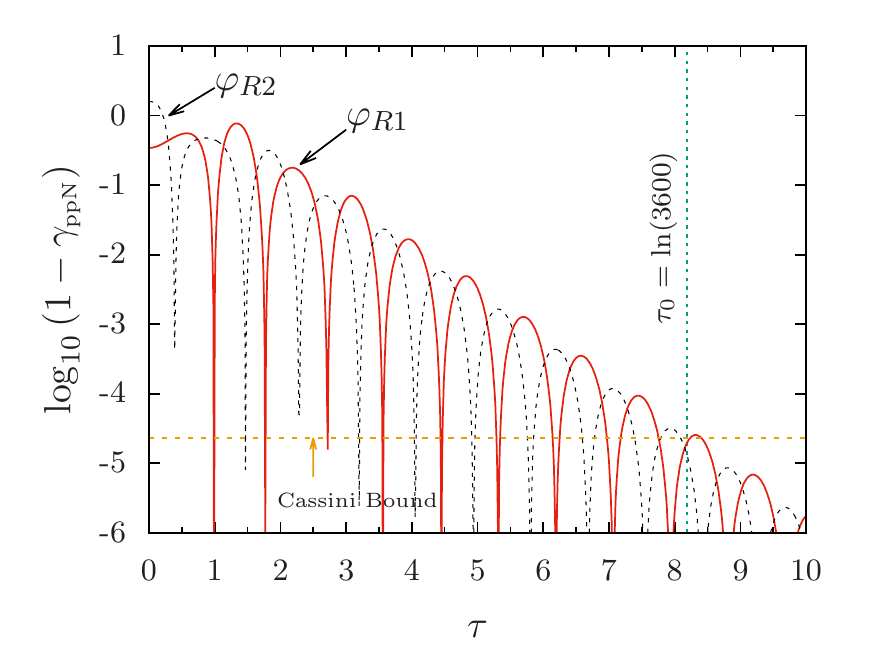}
	\caption{\label{fig:ic5_gammappN_v_time} (Color Online) $|1-\gamma{\ppN}|$ using both initial conditions (red solid for $\varphi_{R1}$ and black dashed for $\varphi_{R2}$) with $\beta = -4.5$ and $\delta = 36$. The horizontal yellow dashed line marks the Cassini bound placed on these theories today, while the vertical line at $\tau_{0}=\ln(3600) \approx 8.2$ corresponds to the present time. Observe that although the solar-system constraint is passed for both initial conditions at $\tau_{0}$, it is not a little $\tau$-time later.}
\end{figure}

One would then think that if the scalar is such as to pass solar-system tests after evolving by $\tau_0 \approx 8.2$, then such tests would also be passed for all later times, but this is not necessarily the case. The reason is that the scalar exhibits oscillatory motion, and thus, it is possible that $\varphi$ is crossing the minimum right at $\tau_0$. This is evident in Fig.~\ref{fig:ic5_gammappN_v_time}, which shows the evolution of $1-\gamma_{\ppN}$ as a function of $\tau$ for a set of $(\beta,\delta)$. At the beginning of the evolution solar-system constraints are clearly not satisfied since the slope of the potential, i.e. $\alpha(\varphi)$, is too large. Observe the oscillations about the minima of the potential signaled by the dips, which represent times at which $\alpha(\varphi)=0$. Observe also that at $\tau = 8.2$ solar-system constraints are satisfied by both initial conditions $\varphi_{R1}$ and $\varphi_{R2}$, but after evolving to later times (e.g.~to $\tau \approx 8.3$) they are not. Notice, however, that although the $\tau$ difference during which tests are not passed seems small ($\Delta \tau \approx 10^{-1}$), this is a very long time interval $\Delta t$.

\begin{figure*}[t]
	\centering
	\includegraphics[width=3.5in]{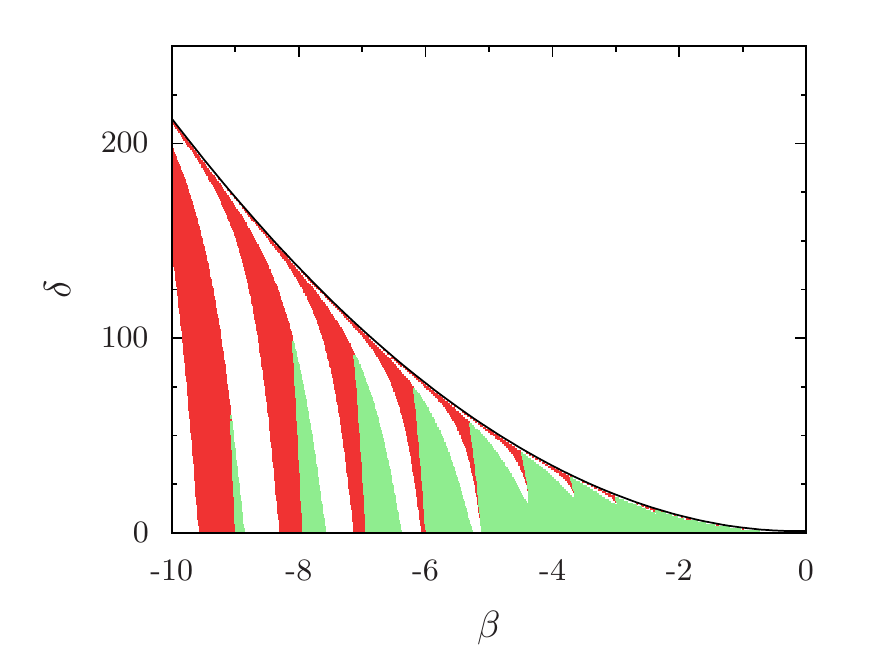}
	\includegraphics[width=3.5in]{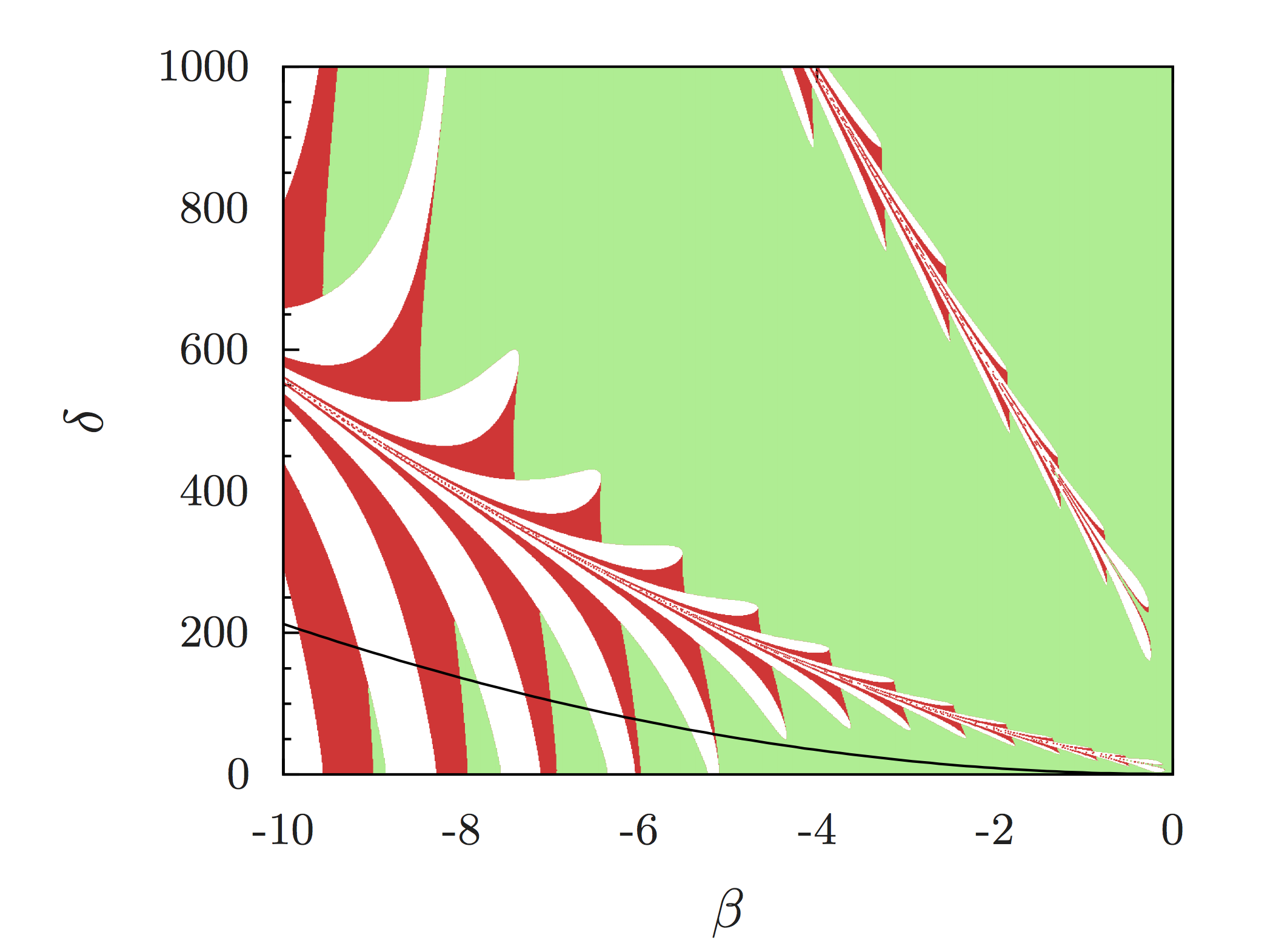}
	\caption{\label{fig:parameter_space_both} (Color Online) 
		Left: $(\beta,\delta)$ parameter space for evolution with initial position $\varphi_{R1}$, the red region corresponding to points that satisfy the Cassini bound today but not for all future times, and the green to points that satisfy it today and all times in the future. The black line through the middle marks the boundary of the regions where $\varphi_{R1}$ exists (below) and does not exist (above).
		Right:  $(\beta,\delta)$ parameter space for evolution with initial position $\varphi_{R2}$ with red/green regions having the same meaning as in the left panel. For reference we also include the black line to link the left and right panels.
	}
\end{figure*}

With no \emph{a priori} knowledge of which initial condition the radiation era leaves the scalar in, we must consider theories for which the evolution of the scalar field with both initial conditions of the previous section leads to passing solar-system constraints. From an analysis of how each of the initial conditions evolves and demanding that $|1-\gamma_{\ppN}| \leq 2.3\times10^{-5}$ for $\tau \geq 8.2$, stringent upper bounds can be placed on the $(\beta,\delta)$ coupling parameter space, as shown in Fig.~\ref{fig:parameter_space_both}. The green regions represent the values of $(\beta,\delta)$ where the cosmological evolution of the scalar field leads to scalar-field values that satisfy the current Cassini bound today and for all future times. Red regions in Fig.~\ref{fig:parameter_space_both} represent the values of $(\beta,\delta)$ that satisfy the Cassini bound today but fail to do so in the future. Empty regions (white) correspond to all other values of $(\beta,\delta)$, i.e.~those that do not satisfy the Cassini bound today and are thus ruled out. The black line running through the plot designates the separation between regions of parameters space where both $\varphi_{R1}$ and $\varphi_{R2}$ exist (below the line) and those where only $\varphi_{R2}$ does (above the line). When considering values of $(\beta,\delta)$ below the line one must consider the intersection (see Fig.~\ref{fig:cosmo__3}) of the two regions as being valid theories such that regardless of the initial condition ($\varphi_{R1}$ or $\varphi_{R2}$) of matter domination, the theory remains consistent with solar-system tests.

\begin{figure}[h]
	\centering
	\includegraphics[width=3.5in]{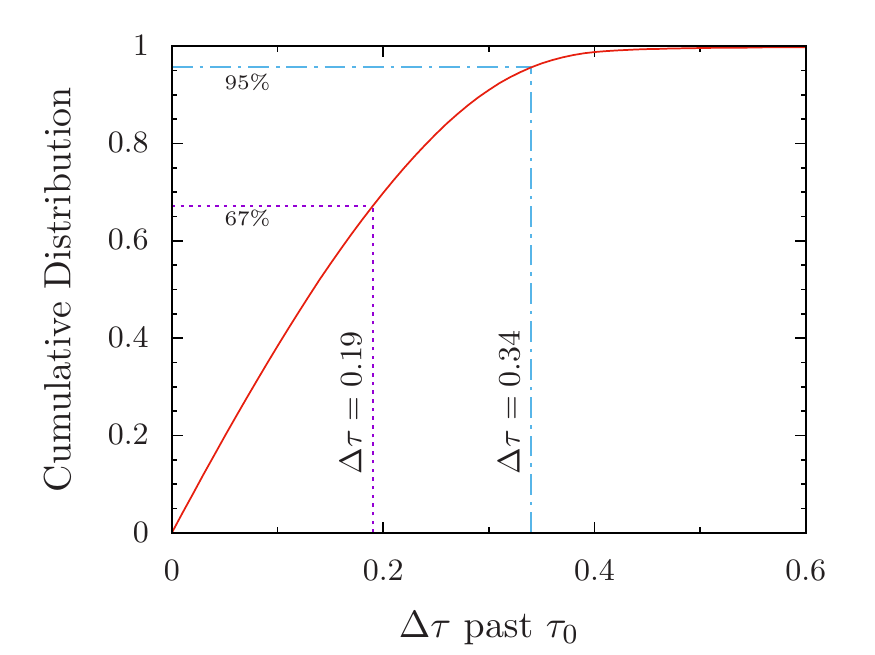}
	\caption{\label{fig:timescales} (Color Online) A cumulative distribution of the points in the $(\beta,\,\delta)$ parameter space that eventually break solar-system tests as a function of how long into the future this occurs. After $\Delta\tau = 0.19$ has passed we find that 67\% of all cases now violate solar-system tests and 95\% of all cases will fail by $\Delta\tau = 0.34$. }
\end{figure}

For the regions in parameter space where solar-system tests are passed today but not in the future (red regions) we can determine a time scale at which the cosmological evolution will becomes inconsistent with future observations. Figure~\ref{fig:timescales} shows a cumulative distribution of how long after today it takes these points in parameter space to violate solar-system tests. We find that 67\% of them fail by a time $\Delta\tau = 0.19$ has passed and 95\% failed after $\Delta\tau = 0.34$. In terms of coordinate time measured in years, i.e. on a human scale, however, these are enormous timescales on the order of $10^9$ years which are in the very distant future. This means that requiring that all future solar-system tests be passed (green regions in Fig.~\ref{fig:parameter_space_both}) may be too conservative, and one may instead only require that tests be passed at least today (the union of green and red regions in Fig.~\ref{fig:parameter_space_both}). The region of parameter space which allows solar-system tests to be passed, at least today, is shown in Fig.~\ref{fig:cosmo__3} for both initial conditions.

The analysis presented above, however, neglects the fact that for sufficiently small $\delta$ (and in particular, in the limit as $\delta \to 0$), solar-system constraints will {\it not} be passed, as the theory reduces to the original DEF theory with $\beta<0$. This is because as $\delta$ becomes very small (relative to $\beta$), the potential becomes deeper and steeper, and it is thus easier for the scalar field to reach the attractor solution during inflation, which we know violates solar-system constraints. Of course, this solution is not reached if the initial conditions for the scalar field are highly fine-tuned close to the minimum of the potential. However, the level of fine-tuning required grows as $\delta$ decreases, making it more and more unlikely that random initial conditions at inflation would lead to a scalar field that satisfies solar-system constraints today. To avoid this fine-tuning problem, we set a minimum value for $\delta$ by requiring that none of the initial conditions on $\varphi$ that fall between the zero crossings of the potential $V_{\alpha}$ (i.e. $|\varphi|= \sqrt{-2 \beta/\delta}$) lead to the attractor solution\footnote{Note that because the Planck mass is factored out in the action of Eq.~\eqref{EinsteinAction}, $\varphi$ is dimensionless, and $|\varphi|$ of order unity (as is the case here) corresponds to Planck-scale excitations of $\varphi$. These are ``natural'' initial conditions at the beginning of inflation, hence our requirement on $\delta$ ensures that the scalar field is unlikely to be trapped in a runaway solution during inflation, i.e.~outcome 1 discussed in Sec.~\ref{sec:inflation} can never take place.}. For the range of $\beta$ we considered, this places a lower-bound on $\delta$ of roughly unity, i.e.~$\delta \gtrsim 1$.

%--------------------------------------------------------------------
\subsubsection{Dark Energy Domination}
To determine the complete evolution of the scalar field one must also consider the dark energy dominated era that follows the matter-dominated one. In this case, the evolution of the scalar field is the same as during inflation [the scalar field evolves as given in Eq.~({\ref{eq:inflation_evo}})], and it will thus continue to be damped to the minimum of the potential. We neglect this era in our study because its effects only become significant for redshifts $Z\lesssim 1$, which corresponds to a small $\Delta \tau \lesssim 0.7$. Because the dynamics of the solutions we find occurs on much larger time scales and because the matter era damps the solutions as well, we expect our conclusions to hold, at least qualitatively, even in the presence of a dark energy dominated era. 

%------------------------------------------------
\subsection{General Coupling Potentials}\label{General_Coupling_Potentials}

We can now use the insights gained from the previous section to understand the evolution of the scalar field in theories with more generic conformal coupling potentials. The key idea to remember about the potential in Eq.~\eqref{eq:coupling_pot_delta} is that it possessed global minima that the scalar could eventually settle to, i.e.~the potential was bounded from below. Because of this feature, the scalar field could damp toward one of these minima and settle down so as to pass solar-system constraints. This idea can be extended to other polynomial forms of $V_{\alpha}$ and $\alpha(\varphi)$.

Let us first consider Eq.~(\ref{eq:gen_alpha}) with the highest power in $\varphi$ even, such that the highest power in $V_{\alpha}$ is odd. Such a potential is not bounded from below at either $\varphi \to +\infty$ or $\varphi \to -\infty$, and thus, it will eventually diverge to negative infinity.\footnote{A potential unbounded from below 
would also be expected to lead to quantum mechanical instabilities as it allows no ground state. This forces unbounded potentials, such as the one considered in footnote 18 of Ref.~\cite{Damour:297525}, to be extremely fine tuned and therefore require a very specific set of initial conditions to pass solar-system tests.}
The BBN constraint in Eq.~(\ref{potential_rad}) still holds for all potentials, and will therefore determine the particle's initial position at the end of the radiation-dominated era. This constraint will always lead to at least one initial condition in the unbounded regime of the potential, leaving the scalar field no choice but to run away toward the attractor solution, rapidly violating solar-system constraints. Therefore, without \emph{a priori} knowledge of the initial conditions at inflation or some argument that eliminates the initial condition that unavoidably leads to an attractor solution, all potentials of this form are immediately ruled out by requiring that initial conditions not be fine-tuned. 

A similar argument also applies to coupling potentials whose highest power is even but with a negative coefficient. These potentials have two regions that approach $-\infty$, and thus, they will result in run-away solutions for the scalar field. By the same initial condition argument discussed above, these theories will not generically pass solar-system tests after cosmological evolution.

The probability that the scalar will find its way to the unbounded part the potential only increases with the inclusion of multiple scalar fields. Of course, the scalar field might evolve in these potentials and never reach these regions, for some initial conditions. However, the only way to guarantee that this does not occur for generic initial conditions is to demand that the potential be bounded from below, and this is the simplest and safest assumption to make.  

Moreover, all of our discussion in Sec. \ref{sec:inflation} and \ref{matter_sec} can be extended to all other polynomial potentials whose highest power is even and has a positive coefficient. These potentials are qualitatively similar to the the quartic one we have considered thus far, in the sense that there exists a global minimum for the scalar field to settle near. Locally, near the minima, these potentials look nearly identical to the one we have considered here and thus one would expect qualitatively similar results. Considering these higher order potential, however, comes at the cost of adding more degrees of freedom and coefficients to constrain, unnecessarily complicating the problem even further.

%%%%%%%%%%%%%%%%%%%%%%%%%%%%%%%%%%%%%%%%%%%%%%%%%%
\section{Neutron Stars and Scalarization}
\label{Neutron Stars and Scalarization}

In this section, we discuss the basics of scalarization and under what conditions it can occur. Note that we focus on spontaneous scalarization in isolated NS's, because one expects theories where the
latter is not possible to not allow for dynamical/induced scalarization in binaries~\cite{Barausse:2012da,Palenzuela:2013hsa}. In more detail, we first consider NS's in the original DEF theory. We then extend these calculations to the modified DEF theory with the cubic conformal coupling function presented in the previous section to show that scalarization cannot occur. We conclude by extending our arguments to more generic potentials. 

\subsection{DEF Theory}
For a spherically symmetric, non-rotating star, we can write the Einstein-frame line element as
\be
ds^2_* = -e^{\nu(r_*)}dt^2_* + \frac{dr^2_*}{1-2\mu(r_*)/r_*} + r^2_*d\Omega^2_*\,\,,
\label{eq:matric_ansatz}
\ee
where $\mu$ and $\nu$ are functions of $r_*$ and are determined from the field equations. The matter inside old and cold NS's can be described through a perfect fluid stress-energy tensor given in Eq.~(\ref{eq:perfect_fluid}).

Using the line element above in Eqs.~(\ref{eq:Einstein_eqns}-\ref{eq:scalar-field-evol}) and applying the  stress-energy conservation condition in the Jordan frame, $\nabla_{\mu}T^{\mu\nu} = 0$, we arrive at the set of first-order differential equations~\cite{Damour:1993hw} 
\ba
	\mu' &=& 4\pi G r_*^2 A^4(\varphi)\rho + \dfrac{1}{2}r_*(r_*-2\mu)\psi^2\,\,,\nn\\
	\nu' &=& r_*\psi^2 + \dfrac{1}{r_*(r_*-2\mu)}\left[2\mu + 8\pi G r_*^3 A^4(\varphi) \rho\right]\,\,,\nn\\
	\varphi' &=& \psi\,\,,\nn\\
	\psi' &=& \dfrac{4\pi G A^4(\varphi)}{(r_*-2\mu)}\left[ \alpha(\varphi) (\rho - 3p) + r_*\psi(\rho - p)\right]\nn\\ &\,&\,\,\,-2\psi\dfrac{(1-\mu/r_*)}{(r_*-2\mu)}	\,\,,\nn\\
	p' &=& -(\rho + p)\left( \nu'/2 + \alpha(\varphi)\psi\right)\,\,.
	\label{eq:NSstructure}
\ea
Note that the density and pressure in these equations are the Jordan-frame ones.
To close the system of equations, we use a simple polytropic EoS in the Jordan frame: 
\begin{gather}
\label{eq:EoS}
p = K \; \bar{\rho}^{\Gamma}\,,\\
\rho=\bar{\rho}+\frac{p}{\Gamma-1}\,,
\label{eq:EoS2}
\end{gather}
with $\bar{\rho}$ the Jordan-frame baryonic density, and with $\Gamma = 2$ and $K = 123\,G^3 M_{\odot}^2$ following Ref.~\cite{Barausse:2012da,lorene}. The choice of EoS affects the mass and radius of the NS, as well as the exact compactness at which spontaneous scalarization occurs. However, within the set of realistic neutron-star EoSs, the particular equation-of-state choice made does \emph{not} affect whether scalarization exists in the first place or not.

%details of the solution
\begin{figure*}[t]
	\centering
	\includegraphics[width=3.5in]{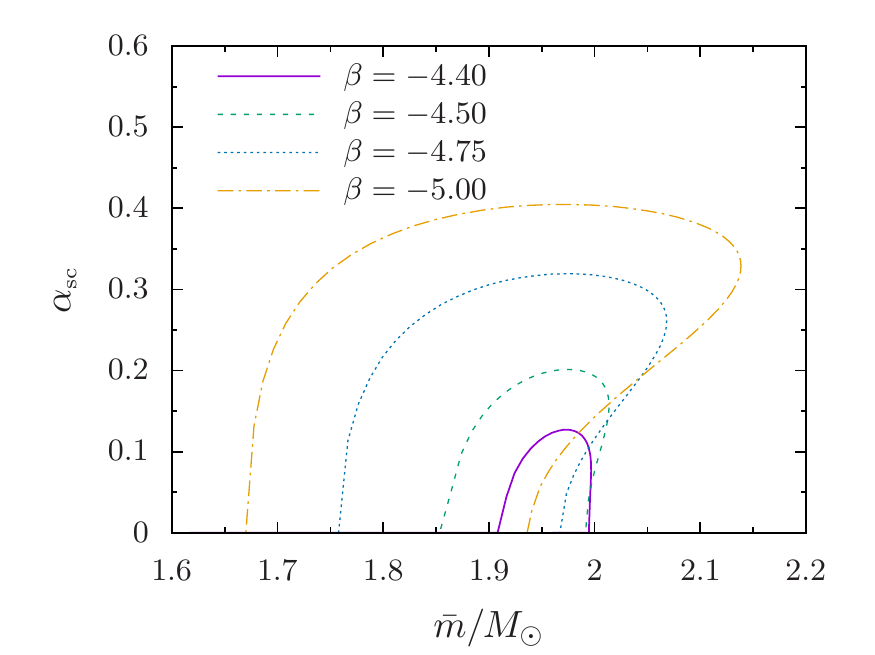}
	\includegraphics[width=3.5in]{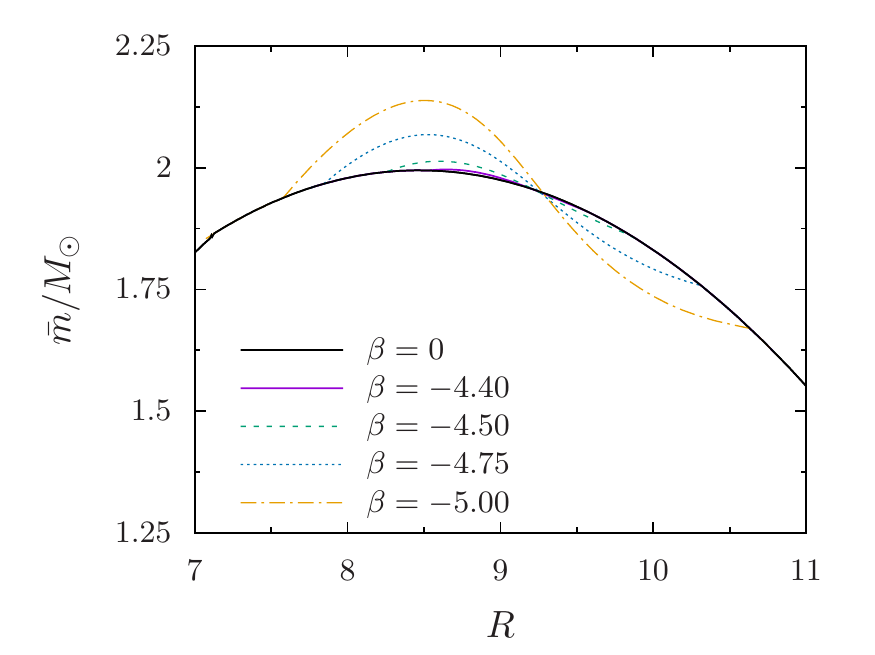}
	\caption{ \label{fig:charge_plot_no_delta} (Color Online) 
		Left: Scalar charge as a function of baryonic mass in DEF theory [$\alpha(\varphi)=\beta\varphi$] with various values of $\beta$. For simplicity, we choose $\varphi_{\infty} = 0$, since this quantity is constrained to be close to zero by solar-system experiments. Observe that the scalar charge activates spontaneously when the mass exceeds a certain critical value, which depends on $\beta$. 
		Right: Mass-radius curves in DEF theory with Jordan-frame quantities $\bar{m}$ and $R$, the same choices of $\beta$, and $\varphi_\infty=0$. The GR curve is shown in black, and one can see that the more negative $\beta$ becomes, the greater the deviation from GR.
	}
\end{figure*}
We numerically solve the system of equations described above for a set of stars parameterized by their central density. In particular, we use {\texttt{Mathematica}}'s
default integrator, which uses an LSODA approach, switching between a non-stiff Adams method and a stiff Gear backward differentiation formula method~\cite{mathematica}.
Inside the star, the pressure is given as a function of the density via the polytropic EoS of Eqs.~\eqref{eq:EoS}--\eqref{eq:EoS2}, while outside the pressure and density vanish. The boundary of the interior and exterior region, i.e.~the radius of the star, is defined by where the pressure vanishes. Our code integrates from the center of the star to an effective spatial infinity, thus fully determining the metric and the scalar field in the entire spacetime.
Near spatial infinity, the scalar field decays as 
\be
\varphi \approx \varphi_{\infty} + \frac{\omega}{r^{*}}\,\,.
\label{eq:phi_vac}
\ee
The quantity $\varphi_{\infty}$ is the asymptotic (at spatial infinity) value of the scalar field (which we fix to a specified value as a boundary condition), while $\omega$ is related to the \emph{scalar charge} of the star $\alpha_{\SC}$ via~\cite{Damour:1993hw}
\be
\alpha_{\SC} = \dfrac{\omega}{G m_*}\,\,,
\ee
where $m_*$ is the Einstein-frame ADM mass of the star; notice that we add a subscript $sc$ to the scalar charge to distinguish it from  the conformal coupling $\alpha$ in Eq.~\eqref{eq:alpha}. We find $\omega$, and thus $\alpha_\SC$, by extracting the $1/r$ part of the scalar field by fitting its exterior solution from our numerical calculations.

Figure~{\ref{fig:charge_plot_no_delta}} shows the results of our numerical calculations for DEF theory, which reproduce old results from the literature~\cite{Damour:1993hw,Damour:1992we}. (We assume here $\varphi_\infty=0$.) One can see that the scalar charge ``spontaneously'' turns on at a critical baryonic mass $\bar{m}_{\rm crit}$. One can also see in Fig.~{\ref{fig:charge_plot_no_delta}} that for $\bar{m}>\bar{m}_{\rm crit}$ 
there are two branches of solutions, one with $\alpha_{\SC}\neq0$ and one with $\alpha_{\SC}=0$. This second branch is unstable to perturbations, i.e. those solutions will either collapse or evolve to the stable branch. The more negative $\beta$ becomes, the larger the maximum scalar charge. These results are quantitatively dependent on the EoS used and for our choice, scalarization occurs only when $\beta \lesssim -4.4$.

%par8: Damours calculation
Let us now provide a physical explanation for why spontaneous scalarization occurs, following the arguments in~\cite{Damour:1993hw}. Consider then the evolution equation of the scalar field in the weak-gravity static limit, such that $\square \rightarrow \delta^{ij} \nabla_{i} \nabla_{j}$. Let us further consider a constant density star (with negligible
pressure, following the weak-field assumption), such that $-4\pi G T^* \rightarrow 4\pi G \rho_{*} = 3 G m R^{-3} = 3\mathcal{C} R^{-2}$, with $\mathcal{C} = Gm/R$ the compactness and $R$ the stellar radius. This leaves us with the simple equation 
\be
	\nabla^2\varphi = \text{sign}(\beta)K^2\varphi\,\,,
\label{eq:simplified_KG_DEF}
\ee
where $K^2 = 3 \; \mathcal{C} \; |\beta| \; R^{-2}$ when $r_*<R$ and $K = 0$ when $r_*>R$. The solution must be regular at the 
center, $\varphi(0) = \varphi_{c} = \text{finite}$ and $\varphi'(0)=0$, and must be continuous and differentiable at the surface $r_{*} = R$. When $\beta<0$, these conditions lead to the interior solution\footnote{Technically, the regularity conditions at the center lead to the interior solution $\varphi_{\rm int} = \varphi_{c} \sin(K r_{*})/(K r_{*})$, while the exterior solution is $\varphi_{\rm ext} = \varphi_{\infty} + \omega/r_{*}$. The matching conditions at the surface, $\varphi_{\rm int}(r_{*} = R) = \varphi_{\rm ext}(r_{*} = R)$ and $\varphi_{\rm int}'(r_{*} = R) = \varphi_{\rm ext}'(r_{*} = R)$, relate the central value of the field to its asymptotic value at spatial infinity $\varphi_{c} = \varphi_{1}/\cos(K R)$.}
\be
	\varphi = \dfrac{\varphi_{\infty}}{\cos(K R)}\dfrac{\sin(K r_{*})}{K r_*}\,\,.
\label{eq:Damour_solution}
\ee
One can see that when $K R = \pi/2$, the scalar field can be amplified inside the star, even when $\varphi_{\infty} \approx 0$. This is how spontaneous scalarization occurs in DEF-like theories.
When $\beta>0$, however, the solution can be obtained  by replacing sin and cos with sinh and cosh respectively in Eq.~(\ref{eq:Damour_solution}). In this case, one typically finds a de-amplification of the scalar field inside the star which suppresses any deviations from GR.

\begin{figure*}[t]
	\centering
	\includegraphics[width=3.5in]{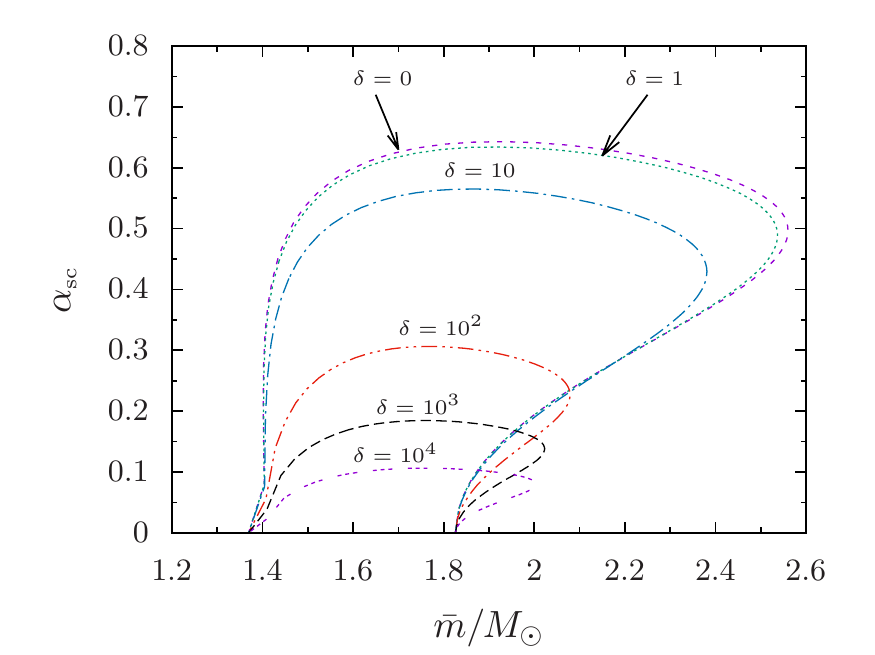}
	\includegraphics[width=3.5in]{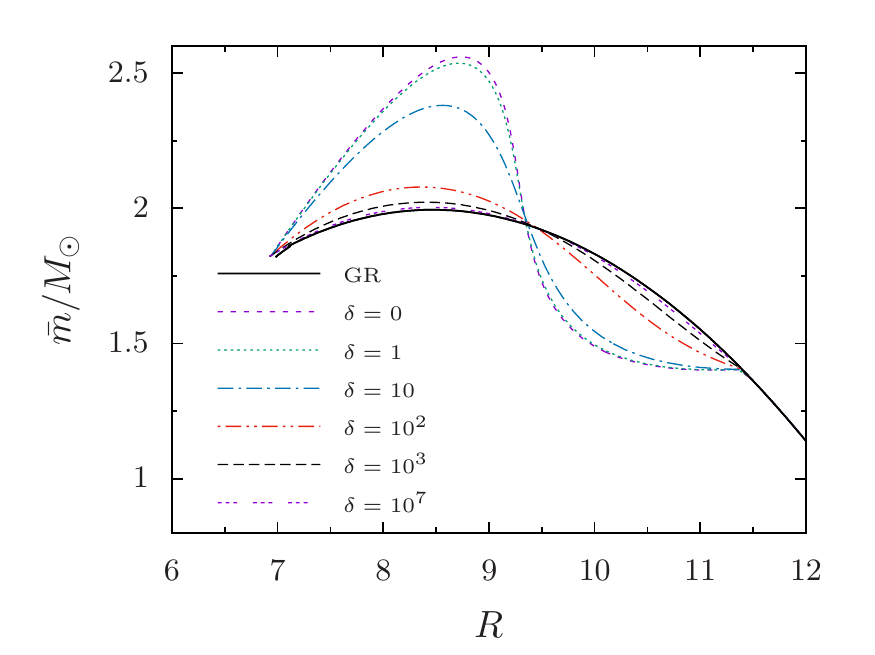}
	\caption{ \label{fig:mDEF_charge_MR} (Color Online) 
		Left: Scalar charge as a function of baryonic mass in modified DEF theory [$\alpha(\varphi)=\beta\varphi + \delta\,\varphi^3$] with $\beta= -6$ and various values of $\delta$. We again choose $\varphi_{\infty} = 0$ to see the effect that the $\delta$ term has on the previous DEF theory results. We see again that the scalar charge activates spontaneously when the mass exceeds a certain critical value, and a larger value of $\delta$ causes the charge to become smaller.
		Right: Mass-radius curves in modified DEF theory with Jordan-frame quantities $\bar{m}$ and $R$, $\beta = -6$, and the same choices of $\delta$. With the black curve representing GR again, we see that 
		the deviations away from GR are maximized when $\delta = 0$ (i.e. when the theory reduces to DEF theory), while larger values of $\delta$ decrease any deviations from GR.
	}
\end{figure*}

A similar argument holds when $\beta >0$ if one considers a NS whose trace of the stress-energy tensor, i.e. $T = -\rho + 3p$, is positive such that the overall sign on the right-hand side of Eq.~(\ref{eq:simplified_KG_DEF}) is still negative. This will lead to the same instability in the star and give a solution similar to Eq.~(\ref{eq:Damour_solution}), in which case the scalar field becomes amplified in the regions where $p>\rho/3$. Indeed, it has been shown that for $\beta$ very large, i.e. $\beta\gtrsim 100$~\cite{Mendes:2016fby}, scalarization can occur in standard DEF theory for certain NS equations of state; note that we here restrict attention to the $\beta<0$ case.

\subsection{Modified DEF Theory}

%par1: for mDEF theory we can make the same assumptions and solve the equations to numerically to extract the properties NSs in these theories
To study NS's in modified DEF theory we must numerically solve the equations of the previous section but with $\alpha(\varphi)$ defined as in Eq.~(\ref{alpha_delta}). As a first pass, we will continue to assume that $\varphi_{\infty} = 0$ to gain insight on how the inclusion of the $\delta$ term affects the results of the previous subsection. The left panel of Fig.~\ref{fig:mDEF_charge_MR} compares the scalar charge present for several orders of magnitude in $\delta$. We see that spontaneous scalarization still occurs and it even ``turns on/off'' at the same values of $\bar{m}$ as in the ($\delta =0$) DEF theory case. One can see that adding the $\delta$ term suppresses the scalar charge of the star and drives the solution to that of GR in the limit $\delta \rightarrow \infty$. These results are also evident in the right panel of Fig.~\ref{fig:mDEF_charge_MR} where mass-radius relations are plotted for a large range of $\delta$ for fixed $\beta$. The largest deviations from GR occurs when $\delta=0$ (i.e. DEF theory) and again it is clear that as $\delta \rightarrow \infty$ the NS solutions reduce to those found in GR.

\begin{figure}[h]
	\centering
	\includegraphics[width=3.5in]{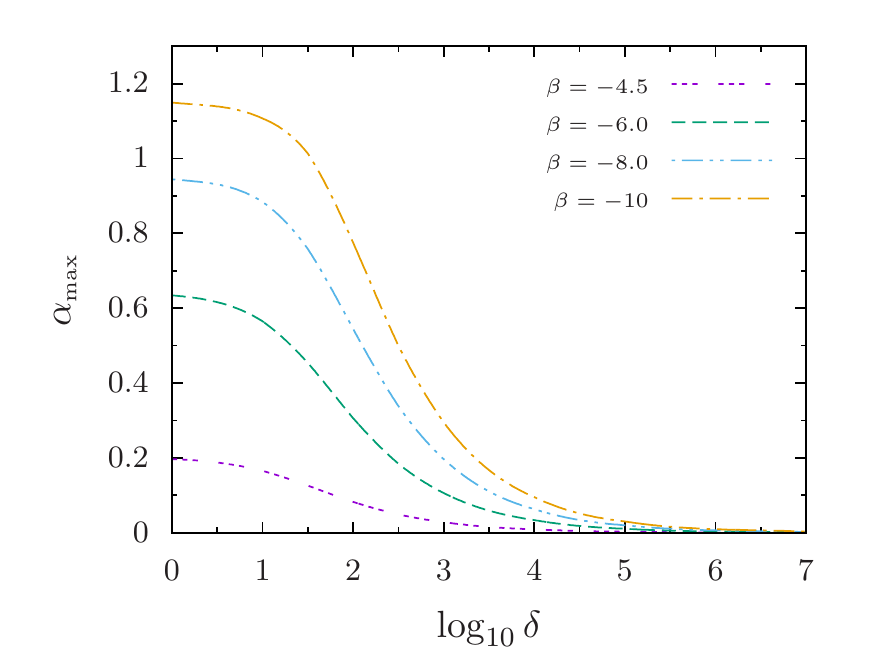}
	\caption{\label{fig:mDEF_max_charge_0} (Color Online)
		Maximum scalar charge as a function of $\delta$ for multiple values of $\beta$ in modified DEF theory. Observe a uniform monotonic decrease as $\delta$ become large for a significant range of negative values of $\beta$.  }
\end{figure}

%par2: max charge and remind the reader that we set the asymptotic value to zero and allowed the field to roll as a results and so is not consistent with our cosmology discussion
One can can go even further and extract the maximum value of the scalar charge as a function of $\delta$ to explore a much broader region of parameter space, comparable to that in Fig.~\ref{fig:cosmo__3}. In Fig.~\ref{fig:mDEF_max_charge_0} we plot this quantity for different choices of fixed $\beta$ . We find a clear monotonic decrease in the maximum scalar charge as $\delta$ increases. Also, not surprisingly, as $\beta$ becomes more negative the (maximum) scalar charge increases, since the curvature of the potential becomes more negative and the potential has a steeper slope, thus allowing the field to become more amplified as a result.

%par3: Qualitatively describe why one would expect there to be no scalar charge since the potential has local positive curvature at the minimum
We are now in a position to link the results from cosmological evolution and solar-system tests to NS's and scalarization. So far, we have assumed $\varphi_{\infty} = 0$ in our NS solutions, but there is a problem with this assumption when we connect to our previous results in Fig.~\ref{fig:parameter_space_both}: $\varphi_{\infty}$ does \emph{not} vanish upon cosmological evolution, but rather it is near the minimum of the conformal potential if it is to satisfy solar-system constraints today. Thus, choosing $\varphi_{\infty} = 0$ \emph{a priori} when building NS solutions (corresponding to the field sitting near the local maximum in Fig.~\ref{fig:initial_conditions}) is completely inconsistent with solar-system observations. Instead, we must set $\varphi_{\infty} \sim \varphi_{\text{min}}$ such that it is near the global minima.

%par4: so what will happen?
Before proceeding numerically, let us take a step back and qualitatively explain what should happen when $\varphi_{\infty} \sim \varphi_{\text{min}}$. The negative curvature of the potential in DEF theory ($V_{\alpha} = \beta\varphi^2/2$) leads to scalarization because the scalar field has the ability to roll in the potential. The same is true in modified DEF theory ($V_{\alpha} = \beta\varphi^2/2 + \delta\,\varphi^4/4$) when $\varphi_{\infty}=0$, because the field sits near the local maximum of the potential and can roll when influenced by matter, which explains our previous numerical results on scalarization. However, if we now set the asymptotic value of the field to be near the minimum of the potential, such that solar-system tests are passed, the field now sits in a region of the potential where the curvature is positive. In regions of local positive curvature we would expect physics to reduce to the case of DEF theory with $\beta >0$, in which case no scalarization occurs because the scalar field can no longer roll. As we will show, our numerical results are consistent with this qualitative idea.
\begin{figure}[h]
	\centering
	\includegraphics[width=3.5in]{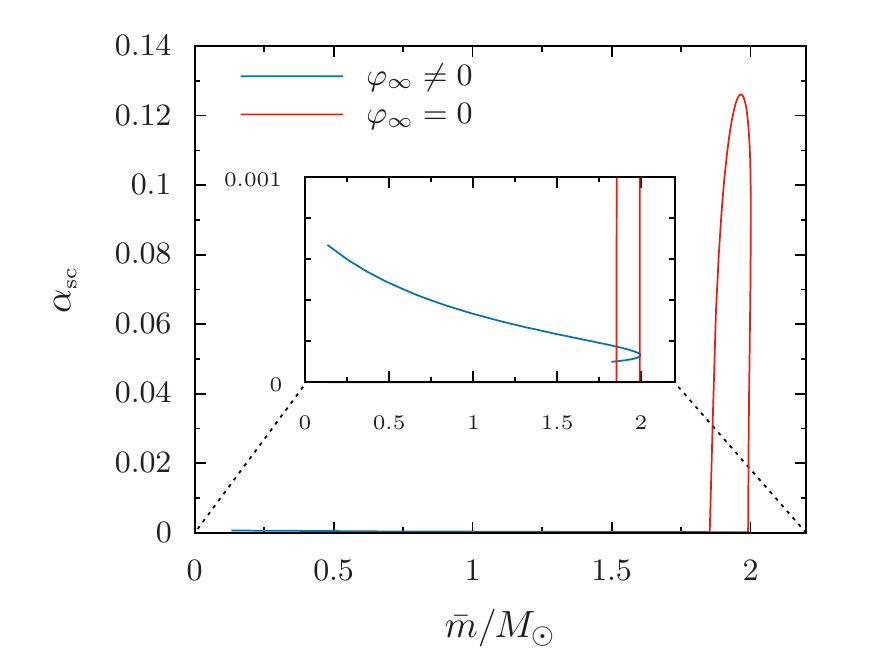}
	\caption{\label{fig:mDEF_charge_near} (Color Online)
		Scalar charge as a function of baryonic mass in modified DEF theory with $\beta = -4.5$ and $\delta=100$. We plot the charge when $\varphi_{\infty} = 0$ and when $\varphi_{\infty}$ lies near (but not exactly at) the minimum of the conformal potential (such that it saturates current solar-system bounds). In the second case (denoted by $\varphi_{\infty} \neq 0$) we see a scalar charge that is always present but never reaches a value greater than $10^{-3}$. This shows that there is no ``spontaneous'' activation of the scalar field in the $\varphi_{\infty} \neq 0$ case.}
\end{figure}
%
%

%par5: show the results and discuss there meaning for when phi_inf = phi_min and conclude by saying that no matter what, when the field is at the minimum we find so scalar charge
Allowing the asymptotic value of the scalar field to be \emph{exactly} at the minimum gives zero scalar charge and all mass-radius curves reduce exactly to GR. This is what one expects, but it may not be the most complete conclusion to make. The Cassini bound requires that the scalar field sit \emph{near} the minimum of the potential, corresponding to the region between the purple dashes in Fig.~\ref{fig:initial_conditions}. The best chance of allowing for scalarization occurs when one saturates this bound and sets $\varphi_{\infty}$ to coincide with one of these dashes near the minimum, giving the scalar field a very small region to roll in. Again, however, we find that spontaneous scalarization does not occur and that all mass-radius curves reduce approximately to that of GR. There does exist a small scalar charge in this case (just like in DEF theory with $\beta>0$ or in JFBD theory), see Fig.~\ref{fig:mDEF_charge_near}, but exactly like in those cases (i) the charge is very small (on the order of $10^{-4}$) and (ii) the charge does not turn on/off suddenly, as one would expect in spontaneous scalarization, but rather it is always present for all masses. 

This behavior is shown in Fig.~\ref{fig:mDEF_charge_near}, which shows the scalar charge as a function of the baryonic mass for two different choices of $\varphi_{\infty}$. In one case, the asymptotic value of the scalar charge $\varphi_{\infty}$ is chosen to be near the minimum of the conformal potential, saturating solar-system constraints (i.e.~$\varphi$ is set equal to one of the dashes near the minimum in Fig.~\ref{fig:initial_conditions}). In the other case, $\varphi_{\infty} = 0$, which allows for spontaneous scalarization, but as discussed earlier, is not consistent with the predictions of the cosmological evolution of the field at the present time. These results prove that if modified DEF theory is to remain consistent with solar-system tests after cosmological evolution, then the spontaneous scalarization of NS's is not a phenomenon that can occur.

Does this inconsistency between scalarization and solar-system tests persist for other forms of the conformal coupling potential? Previously we have argued that the only way solar-system tests can be passed (without a mass term in the scalar field action) is if the potential contains a minimum, preferably a global one to prevent the scalar field from diverging (see Sec.~\ref{General_Coupling_Potentials}). Regardless of the exact form of the potential, however, one must require the the scalar field be near the minimum today in order to pass solar-system tests. This requirement then reduces the problem to a local analysis of the potential near the minimum, where the curvature is positive, thus making the analysis for NS's qualitatively similar to that of the potential that we have studied above. Thus, because the scalar field must sit near the minimum of the conformal potential today, the results of this section generalize to any polynomial coupling potential that passes solar-system tests upon cosmological evolution. 

%------------------------------------------------------------------------------------------------
\section{Conclusion}\label{Conclusion}

%par1:
In this paper, we studied scalar-tensor theories of gravity and their cosmological evolution to determine whether such theories are able to pass solar-system tests today while still allowing for scalarization. As expected~\cite{Sampson:2014qqa,PhysRevLett.70.2217,PhysRevD.48.3436}, the theory proposed by Damour and Esposito-Far\`ese does not pass these tests when $\beta<0$, precisely the values of $\beta$ that lead to spontaneous scalarization in strongly self-gravitating systems like NS's. This is because when $\beta < 0$, an attractor basin arises leading to a run-away scalar field solution that violates solar-system constraints today.

%par2:
We have studied a generic modification to DEF theory by considering a conformal coupling function composed of a higher-order polynomial in the scalar field, such that the associated coupling potential is bounded from below. We show that this modification allows the theory to pass solar-system tests today upon cosmological evolution for a wide range of initial conditions. Any potential that \emph{is not} bounded from below allows the scalar field to reach a runaway attractor solution (at least for some initial conditions), and thus, can never pass solar-system tests for generic initial conditions.  Potentials that are bounded from below and pass solar-system constraints, however, do not allow for spontaneous scalarization. This is because cosmological evolution drives the scalar field to the minimum of the conformal potential, thus eliminating the ability of the field to roll when solving for NS configurations. 

%par3: The results that we have arrived at in this paper prove that scalar-tensor thoeires of gravity are indeed a valid thoeries to test GR against.
The results of this paper suggest that if one wishes to construct scalar-tensor theories that can simultaneously pass solar-system constraints and allow for spontaneous scalarization, then modifications to the conformal coupling of this form are not enough. One possibility is to consider the addition of a mass for the scalar field. Indeed, Ref.~\cite{Pretorius:2016wp} has already shown that massive DEF theory still allows for spontaneous scalarization for a very light scalar (masses between $10^{-15}$ and $10^{-9}$ eV). Such massive scalar tensor theories could potentially pass solar-system constraints upon cosmological evolution, an investigation that is currently ongoing. Another possibility is to consider other, non-polynomial, functional forms for the coupling potential, such as that studied recently in~\cite{Mendes:2016fby}. Indeed, the latter reference has recently claimed that such a potential allows for spontaneous scalarization. Whether this model also passes solar-system constraints will also be analyzed in a forthcoming publication.  

\acknowledgements
We thank Gilles Esposito-Far\`{e}se for insightful conversations on scalarization and scalar-tensor theories, and Norbert Wex and Thibaut Arnoulx de Pirey for reading our manuscript carefully and providing useful suggestions. N.~Y.~acknowledges support from the NSF CAREER Grant PHY-1250636. E.~B.~acknowledges support from the European Union's Seventh Framework Programme (FP7/PEOPLE-2011-CIG) through the Marie Curie Career Integration Grant GALFORMBHS PCIG11-GA-2012-321608. This project has received funding from the European Union's Horizon 2020 research and innovation programme under the Marie Sklodowska-Curie grant agreement No 690904.

%%%%%%%%%%%%%%%%%%%%%%%%%%%%%%%%%%%%%%%%%%%%%%%%
\bibliography{master}
\end{document}